\documentstyle[epsf]{aipproc}

\def\eps@scaling{.95}
\def\epsscale#1{\gdef\eps@scaling{#1}}

\def\plotone#1{\centering \leavevmode
    \epsfxsize=\eps@scaling\columnwidth \epsfbox{#1}}

\begin{document}
\title{UV Emission from Elliptical Galaxies Near and Far}

\author{Henry C. Ferguson$^*$ and Thomas M. Brown$^\dagger$}

\address{$^*$ Space Telescope Science Institute,
	      3700 San Martin Drive, Baltimore, MD 21218\\
	$^\dagger$ NASA/GSFC, Code 681, Greenbelt, MD 20771}

\maketitle

\begin{abstract}
        The far-ultraviolet is the most rapidly evolving portion
of the spectrum in both very young galaxies and very old galaxies.
The ``UV upturn'' in the spectra of elliptical galaxies shortward of
2000\AA\ offers a promising probe of the ages and chemical
evolution of very old galaxies. In early-type non-active galaxies
with the bluest $1550-V$ colors, the bulk of the emission arises
from Extreme Horizontal Branch (EHB) stars, along their evolution from
the zero-age HB to the white-dwarf cooling curve. The strength
of the UV-upturn is governed by the fraction of stars that evolve
through the EHB phase, which is in turn governed by age, metallicity,
helium abundance, and other parameters such as stellar rotation
and binarity that might influence the amount of mass loss on the
RGB. Spectral constraints on the nature of the hot stellar population
from Astro-2 are reviewed, and new imaging results from the HST
Faint Object Camera are presented. Attempts to measure evolution through
observations of high-redshift elliptical galaxies in the rest-frame 
UV are reviewed.

\end{abstract}

\section*{Background}

Giant elliptical galaxies show a large variation in the ratios of their
far-UV to optical fluxes. Shortward of 2000{\AA}, most ellipticals have
spectra that rise in $f_\lambda$ toward shorter wavelengths.  This hot
component has been known since the early days of space astronomy
\cite{CW79}. Observations and theory now seem to be converging on a
consensus that the dominant component in UV bright galaxies is extreme
horizontal branch (EHB) stars and their evolutionary progeny
\cite{GR90,FD93,DOR95,BCF94,BFD95,BFDD97,YDK97}.  This conclusion stems from the rather cool
temperature (25000 K) derived for the dominant component in NGC1399
\cite{Ferg91L}, and from computations that indicate that EHB stars can
provide enough far-UV photons over their lifetimes to produce the
elliptical galaxy fluxes, while other candidates such as PAGB stars
cannot.

While it seems clear that EHB stars provide the far-UV flux, it is less
clear how they got there. The general trend observed for globular
clusters is that the horizontal branch (HB) becomes redder with
increasing metallicity. Elliptical galaxies are even more metal rich
than Galactic globular clusters; they must somehow be able to buck the
trend. The HB morphology depends on age, metallicity, helium abundance,
and the amount of mass loss on the red giant branch. The helium-burning
core in HB stars has a mass (0.5 $M_\odot$) that is nearly independent
of these parameters. The position of stars along the HB thus depends on
the envelope mass, which in turn depends on the main-sequence mass and
the amount of mass lost during the RGB phase.  EHB stars (those with
$T_{eff} > 20000$~K) have envelope masses less than 0.05$M_\odot$.
Hence they must arise from stars that have lost nearly all the mass it
was possible for them to lose and still ignite helium in their cores.

There are several plausible ways to produce a minority population of
EHB stars in elliptical galaxies.
 
First, the giant elliptical galaxies may in general be {\it older} than
the galactic globular clusters.  This argument is a natural extension
of the interpretation of the second parameter effect in globular
clusters as being due to variations in age \cite{Lee94,PL95}.  The EHB stars in this model represent the extreme metal-poor
tail of the metallicity distribution, and are 2-4 Gyr older than the
most-metal poor globular clusters. They show up in giant elliptical
galaxies, which are on average metal rich, because these galaxies
formed first, and hence have the oldest stars.

Second, elliptical galaxies may have high helium abundance $Y$
\cite{GR90,BCF94,YADO95}.  At fixed age and metallicity, the
main-sequence lifetime decreases with increasing $Y$. Observations of
nearby star-forming galaxies and the galactic bulge hint at a rather
steep relation ($\Delta Y / \Delta Z > 2$) between helium abundance and
metallicity \cite{Pagel89p201,Renzini94}.  If this is the case, then
old metal-rich populations may have EHB stars, even with standard RGB
mass-loss rates. The required ages ($>7$ Gyr) are not as extreme as in
the Lee model. In these models, the EHB stars arise from the extreme
high-metallicity tail of the abundance distribution.

Third, some other process may act to increase mass loss in a small
fraction of the population. For example, EHB stars could arise from
stars in close binary systems that have shed their envelopes during
interactions with their companions, or they could arise only from
stars with high rotation rates. Such mechanisms could produce EHB
populations from anywhere in the abundance distribution, but might be
enhanced in giant ellipticals through some secondary effect (for
example binary fraction might somehow depend on galaxy metallicity or
velocity dispersion). In this case, the EHB stars may come closer to
reflecting the mean metallicity of the stellar population.

Whatever the mechanism for producing them, the fraction of EHB stars is
likely to be sensitive to age, because the main-sequence turnoff mass
is one of the parameters that influences the mass on the horizontal
branch. Indeed, it is likely that the fraction of EHB stars is {\it the
most sensitive} indicator of age for stellar populations older than
$\sim 5$ Gyr. For solar metallicity and ages 5-18 Gyr, the turnoff mass
varies as $-0.63 \log({\rm age})$, according to the Padova isochrones.
A difference of 0.05 $M_\odot$ is enough to move a star from the
UV-weak to the UV-strong portion of the horizontal branch.  In a
population with fixed age and chemical abundance and where no other
parameters influence horizontal-branch mass, a change in age of only
10\% could introduce changes in the UV/optical flux ratio of more than
an order of magnitude.

The problem in using the UV upturn in elliptical galaxies as an 
age indicator is that it lacks a calibration, either empirical
or theoretical. Clearly we need a better understanding of the 
EHB star population and the mechanisms that produce it. 

What follows is a summary of attempts to constrain the metallicity
and helium abundance of the UV-emitting population from Hopkins
Ultraviolet Telescope (HUT) spectra, a preliminary discussion of the
properties of far-UV point sources seen in new images of the centers
of M31 and M32, and a summary of attempts to detect evolution
in the far UV emission from ellipticals as a function of redshift.

\section*{Astro-2 Spectroscopy}

During the Astro-2 shuttle mission in March 1995, HUT 
\cite{DKLBBDDFFHKKMV92,Ketal95} observed six
elliptical galaxies, obtaining spectra at a resolution of 2-4\AA\
over the wavelengths from 912-1840\AA. The 
$10^{\prime\prime} \times 56^{\prime\prime}$ 
slit covered the central regions of each galaxy. The results
of the observations and modeling are described in detail
by Brown and collaborators\cite{BFD95,BFDD97}.

\begin{figure}
\epsscale{0.9}
\centerline{\plotone{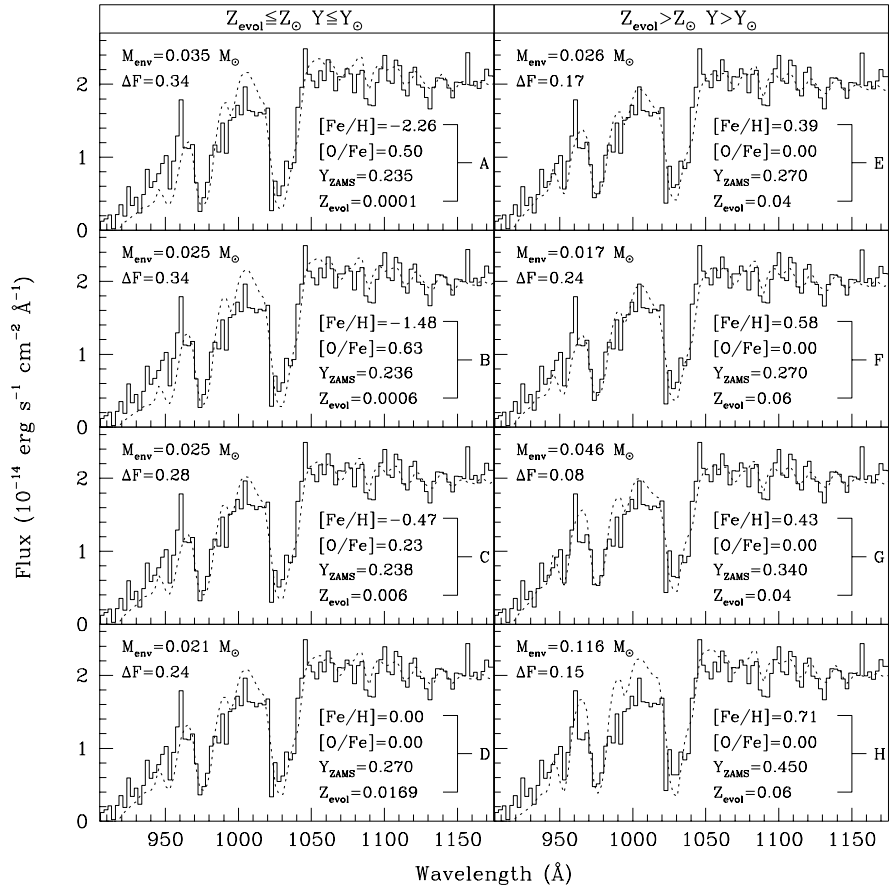}}
\caption{The best-fit PAGB+EHB star evolutionary model from each of the
Dorman et al. (1993) abundance subsets. The HUT spectrum of NGC 4649
(M60) is shown as a solid histogram; the models are shown as dashed lines.
The chemical abundances and envelope mass $M_{\rm env}$ are shown
for each track; the atmospheric abundance is $0.1 Z_\odot$.
The parameter $\Delta F$ is the fractional flux deficit
in the models relative to the data over the wavelength interval
912-970{\AA}. 
Models with subsolar evolutionary abundances (shown in the panels on
the left) signficantly underpredict the observed flux shortward of
970{\AA}.
}
\end{figure}

To model the observations, we have constructed synthetic FUV
spectra for old stellar populations by integrating individual
synthetic stellar spectra over stellar evolutionary tracks.
The Dorman et al. \cite{DRO93} evolutionary models were used
for  EHB stars and their progeny (AGB-Manq\'e
and PEAGB stars).
A new set of model atmospheres of hot stars were constructed
for this purpose\cite{BFD96}. Composite spectra were
constructed for each of over one hundred evolutionary tracks. 

Diffusion processes in the outer layers of hot, high-gravity stars
create inconsistencies between the abundances that determine the
evolution of the stars and the abundances in outer layers of the
stellar atmospheres \cite{MBWF85,LWF87,BWMF88}.
For this reason, we have constructed synthetic
spectra with three different atmospheric metallicities ($Z_{\rm atm} =
Z_\odot, Z_{\rm atm} = 0.1 Z_\odot$, and $Z_{\rm atm} = 0.01_\odot$) for
all of the different evolutionary paths, regardless of the underlying
abundance.

The galaxies have each been fit with single-mass populations of EHB
stars, and with composite populations of EHB stars + PAGB stars.  The
composite population models are more consistent with the data. 
The best fits have
small fraction of the light coming from PAGB stars (less than 20\% at
1400{\AA}), but nevertheless a large fraction (more than 92\%) of the
stars evolve through the PAGB phase.  Figure 1. shows the results for
NGC 4649 for models of different metallicity and helium abundance. The
evolutionary tracks with high helium abundance and high metallicity
(shown in the right-hand panels)
produce the best fits. The low-metallicity alternatives do not produce
enough flux near the Lyman limit.

\begin{figure}
\epsscale{0.9}
\centerline{\plotone{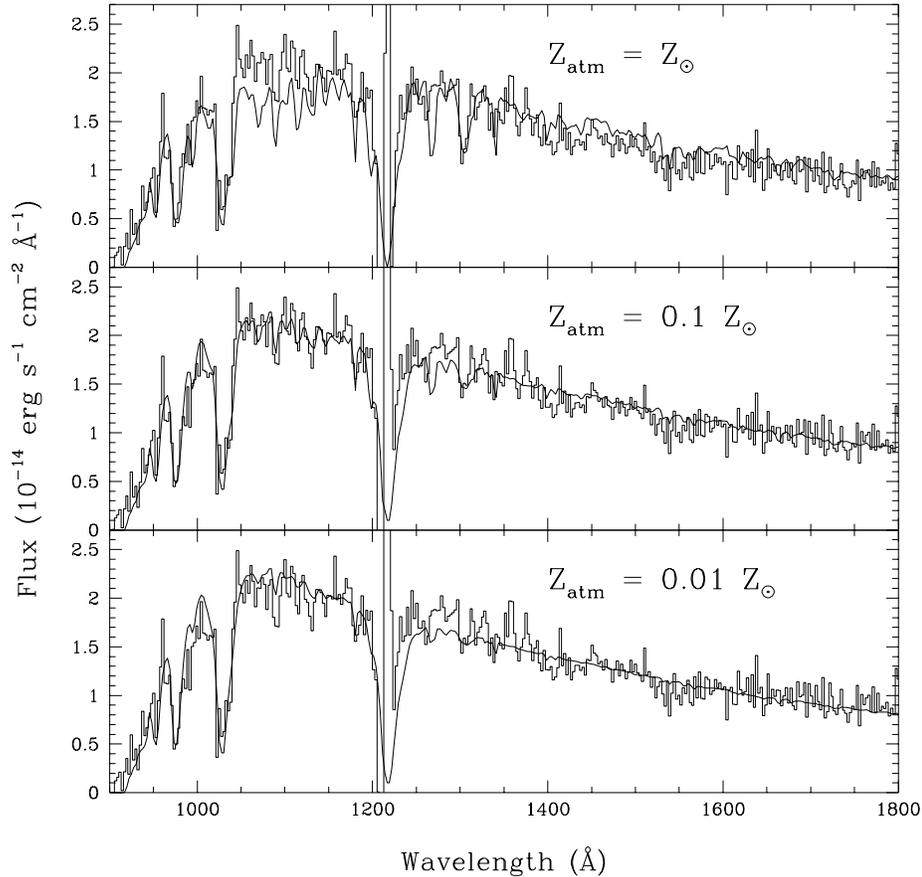}}
\caption{The HUT data for NGC 4649 (histogram) are plotted with the best fit 
EHB$ + $PAGB model (curve). The best-fit model has supersolar evolutionary 
abundances ($Z = 0.04$, $Y = 0.27$), but we have synthesized the spectrum
with model atmospheres of three different abundances. The center
panel, with $Z_{\rm atm} = 0.1 Z_\odot$ provides the best match.
Small differences in the data shown in each panel are due to variations
in the airglow subtraction, since the airglow line fluxes were free
parameters in the fits.
}
\end{figure}

The best-fit models have stellar-atmosphere abundances 
$Z_{\rm atm} = 0.1 Z_\odot$. The preference for this abundance over
solar or 0.01 solar can be seen quite clearly in Fig. 2. However, 
because roughly half the flux comes from stars in the gravity$+$temperature
regime where a complicated balance between radiative levitation, 
gravitational settling, and stellar winds determines the line strengths,
we do not believe the result rules out either the metal-rich or
the metal poor scenario. Among the six galaxies observed on Astro-2,
there is tendency for galaxies with stronger UV upturns to have 
stronger lines, as expected in the high-metallicity scenario\cite{BFDD97}.

\section*{FOC Imaging}

To provide another view of the far-UV emitting population, 
the nearby galaxies M31 and M32 were observed with the HST Faint
Object Camera. The images were taken through the F175W and F275W filters
(the numbers correspond to the filter central wavelengths in nm).
Figures 3 and 4 show color images constructed from the two FOC bands and
a WFPC-2 F555W band image. Preliminary color magnitude diagrams are 
presented by Brown et al. (this volume). There a several inconsistencies
in the photometry that must be cleared up before trusting detailed
comparisons to the theoretical evolutionary tracks. Nevertheless,
broadly speaking the stars populate the portion of the CMD expected
from the spectral analysis.

Above the 6$\sigma$ detection
limit 1349 sources are found in M31 and only 183 in M32. These resolved 
sources account for only a small portion of the far-UV flux in each
galaxy. In M31 the resolved sources produce
39\% of the flux at 1700{\AA}, while in M32 they account for only 8\%. Both
numbers are a bit uncertain due to the photometric errors, and because
some scaling had to be applied to convert from the IUE aperture to the
FOC aperture (assuming the UV surface brightness profile traces the 
optical profile, which may not be a very good assumption). Note, however,
that this calculation is insensitive to red-leak  because the sources
themselves are UV bright and the estimate for the total UV flux comes 
from IUE which is not affected by redleak (or red scattered light).

The fraction of the UV emission that is resolved in M31 is in keeping
with expectations, because much of the far-UV flux arises from stars
near the zero-age HB, which is well below the detection limit of the
images. For M32, the small fraction of the flux that is resolved is a
bit of a surprise, although it was hinted at by earlier FOC
observations\cite{KDABBBCDJKMMPWBGJNSS92,bertola95}. The implication is that most
of the far-UV emission from M32 is {\it not} due to low-mass PAGB
stars.  In a population of pure PAGB stars less massive than $0.7
M_\odot$ at least 15\% of the far-UV flux comes from stars with
temperatures $10^4 < T < 10^5 $K and luminosities $\log L/L_\odot >
3$.  Such stars would be easily detected in the FOC images. For typical
PN central stars of mass $0.6 M_\odot$ the contribution from resolved
sources should be more than 50\%.  Most of the far-UV emission in M32
must come from another source. One possibility is that the diffuse
light comes from normal blue HB stars in the metal-poor tail of
the metallicity distribution. Such emission should be present in all
elliptical galaxies, but is usually overwhelmed by the UV-upturn 
component. 

\begin{figure}
\epsscale{0.7}
\centerline{\plotone{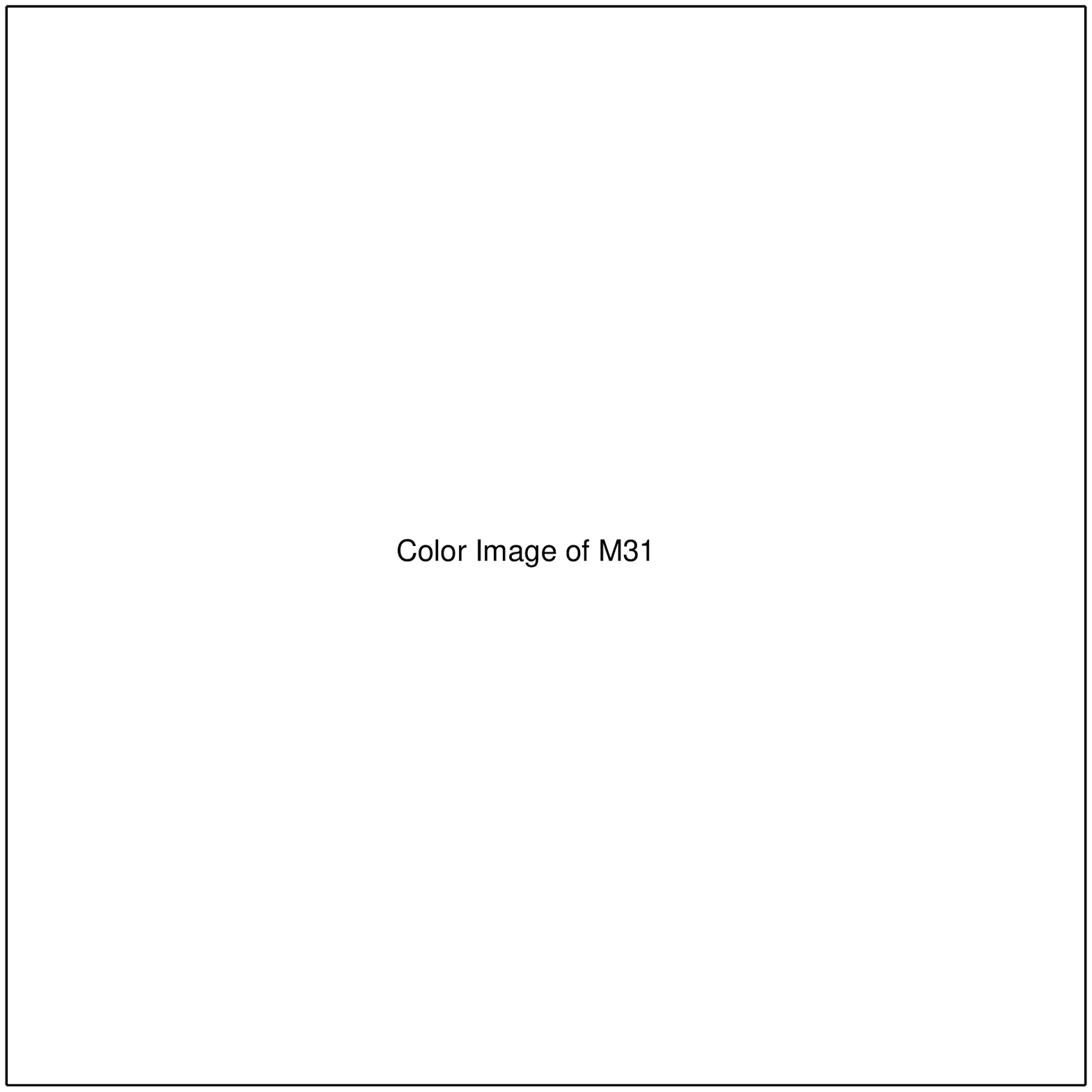}}
\caption{Color image of M31 constructed from FOC F175W and F275W
images and a WFPC-2 PC image taken with the F555W filter.  The field of view
is $14 \times 14$ arcsec. More than 1000 UV-bright sources are detected.
}
\vspace{5pt}
\epsscale{0.7}
\centerline{\plotone{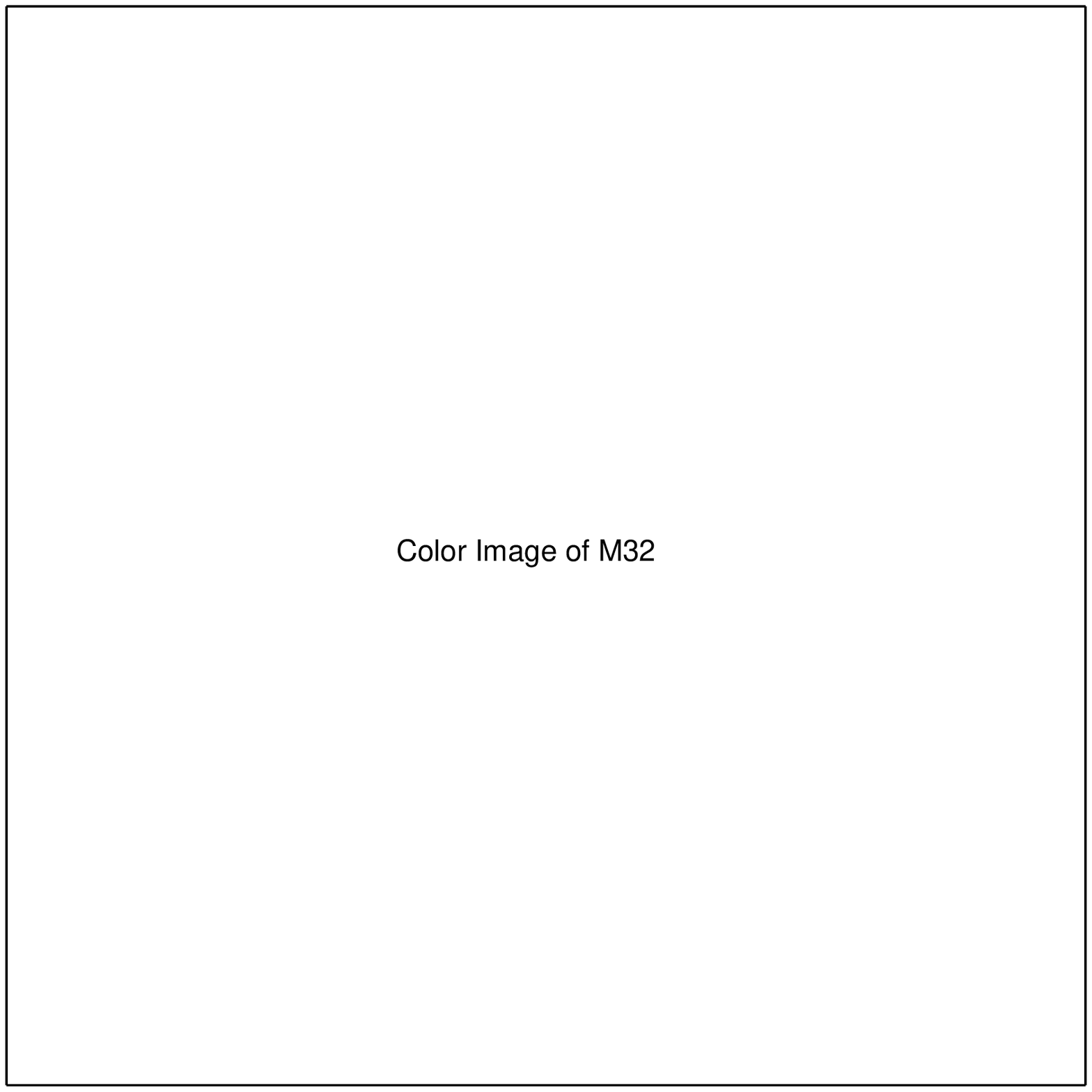}}
\caption{FOC + WFPC-2 Image of the center of M32. The field of view
is the same as for M31. 
}
\end{figure}

\section*{Evolution with Redshift}

\begin{figure}
\epsscale{0.8}
\centerline{\plotone{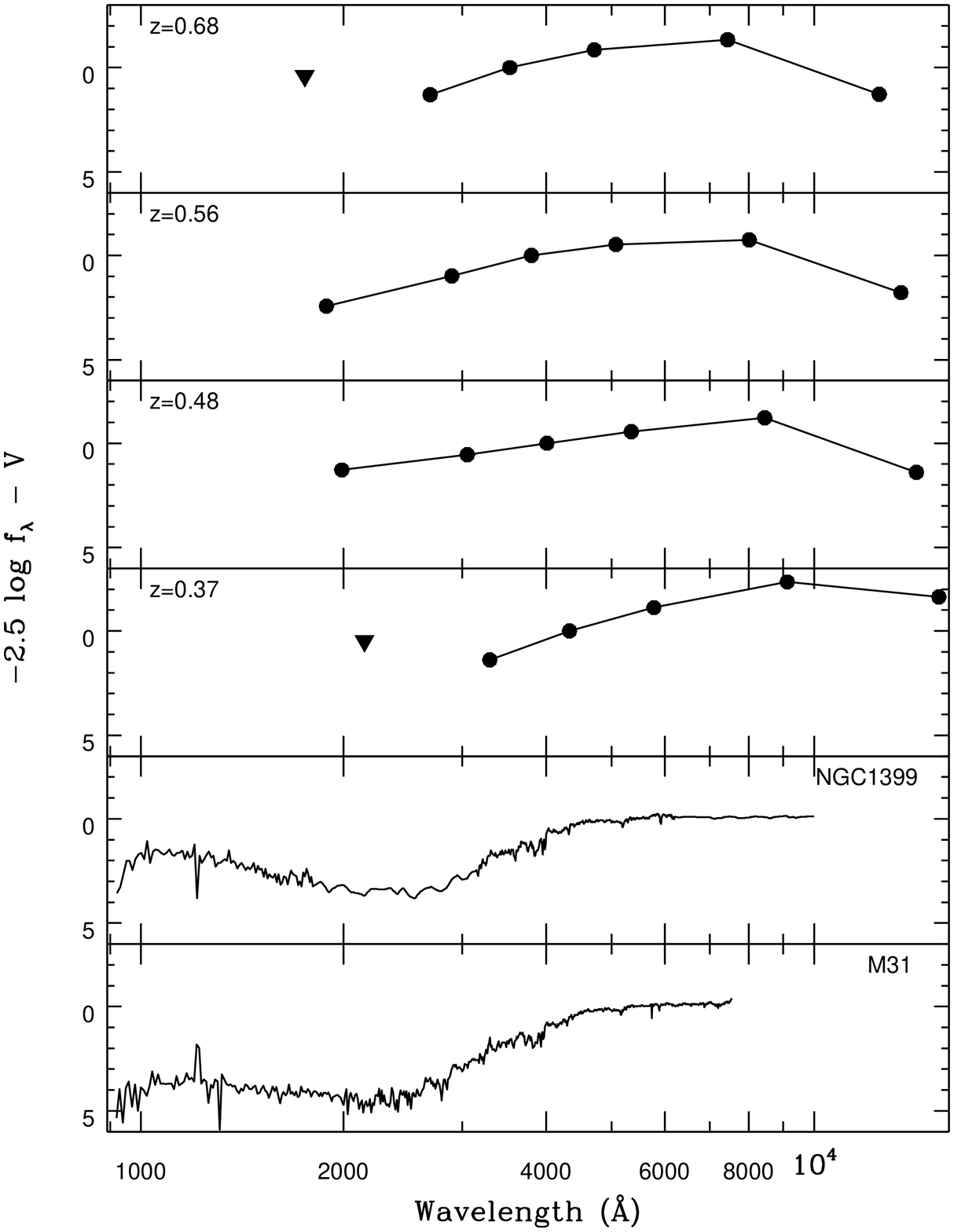}}
\caption{Evolution of the UV upturn. Four elliptical galaxies from the
Hubble Deep field with measured redshifts are shown, together with
composite optical and UV spectra of NGC 1399 and M31 (the optical
and NUV portion of the M31 spectrum is courtesy of Daniela Calzetti).
}
\end{figure}

An obvious test of our understanding of the UV upturn phenomenon
is to try to detect the very strong evolution predicted by theory.
In either the metal-rich or the metal-poor scenario, the UV 
upturn should disappear at lookback times of only 2-5 Gyr. 
By $z > 0.3$, there should be no elliptical galaxies with optical
spectra consistent with passive evolution and strong UV upturns.

Numerous attempts have been made to confirm this prediction using HST.
To date all such attempts have been foiled by either poorer than
predicted sensitivities or poor understanding of scattered light. No
galaxies have been detected (save for Ly$\alpha$ emission from a few
radio galaxies), but the upper limits do not constrain the theory.

The Hubble Deep field provides a new data set to attempt to detect
this evolution. Redshifts have been measured now for over 100 galaxies
in the image, of which about eight appear morphologically to be 
ellipticals.  Figure 5 shows the spectral energy 
distributions derived from the HDF images and the 
Hawaii \cite{SC97} IR images for galaxies with
$0.3 < z < 0.7$. Of the four galaxies, two are detected at rest-frame
wavelengths below 2000{\AA}, and two have only upper linits. 
For the two detected, it will be important to get data at shorter
wavelengths (possible with STIS), and get high S/N spectra in 
the optical to assess whether the spectrum is consistent
with pure passive evolution. 

It will be difficult to get a definitive answer from a few random field
galaxies, since it is possible for a short burst of star formation to
produce a UV upturn without significantly affecting the optical
spectrum (although one might imagine detecting [OII] and/or H$\alpha$
emission in this case). A better measure of evolution  will come from
observations of several elliptical galaxies in each of several clusters
at different redshifts. We can hope to see such observations emerging
from STIS in the next few years.

\subsection*{Acknowledgements}
This work resulted from several different collaborations with Arthur
Davidsen and the HUT team, Ben Dorman, Adam Stanford, Ivan King 
and Jean-Michelle Deharveng, and the STScI HDF working group. 
This work is based in part on observations with the NASA/ESA 
Hubble Space Telescope obtained at STScI, which is operated by
Aura, Inc, under NASA contract NAS5-26555.

\bibliography{apjmnemonic,bib}

\begin{thebibliography}{10}

\bibitem{CW79}
{Code}, A.~D. and {Welch}, G.~A., 1979, ApJ{, 228}, 95.

\bibitem{GR90}
{Greggio}, L. and {Renzini}, A., 1990, ApJ{, 364}, 35.

\bibitem{FD93}
{Ferguson}, H.~C. and {Davidsen}, A.~F., 1993, ApJ{, 408}, 92.

\bibitem{DOR95}
{Dorman}, B., {O'Connell}, R.~W., and {Rood}, R.~T., 1995, ApJ{, 442}, 105.

\bibitem{BCF94}
{Bressan}, A., {Chiosi}, C., and {Fagotto}, F., 1994, ApJS{, 94}, 63.

\bibitem{BFD95}
{Brown}, T.~M., {Ferguson}, H.~C., and {Davidsen}, A.~F., 1995, ApJ{, 454}, L
  15.

\bibitem{BFDD97}
{Brown}, T.~M., {Ferguson}, H.~C., {Davidsen}, A.~F., and {Dorman}, B., 1997,
  ApJ{, 482}, 685.

\bibitem{YDK97}
{Yi}, S., {Demarque}, P., and {Kim}, Y.-C., 1997, ApJ{, 482}, 677.

\bibitem{Ferg91L}
{Ferguson}, H.~C., {Davidsen}, A.~F., {Kriss}, G.~A., {Blair}, W.~P., {Bowers},
  C.~W., {Dixon}, W.~V., {Durrance}, S.~T., {Feldman}, P.~D., {Henry}, R.~C.,
  {Kimble}, R.~A., {Kruk}, J.~W., {Long}, K.~S., {Moos}, H.~W., and {Vancura},
  O., 1991, ApJ{, 382}, L69.

\bibitem{Lee94}
{Lee}, Y.-W., 1994, ApJ{, 430}, L113.

\bibitem{PL95}
{Park}, J.-H. and {Lee}, Y.-W., 1995, preprint.

\bibitem{YADO95}
{Yi}, S., {Afshari}, E., {Demarque}, P., and {Oemler}, A., 1995, ApJ{, 453},
  L69.

\bibitem{Pagel89p201}
{Pagel}, B. E.~J., 1989, In {\em Evolutionary Phenomena in Galaxies,
  }{Beckman}, J. and {Pagel}, B. E.~J., editors,  201 (Cambridge Univ. Press,
  Cambridge).

\bibitem{Renzini94}
{Renzini}, A., 1994, A\&A{, 285}, L5.

\bibitem{DKLBBDDFFHKKMV92}
{Davidsen}, A.~F., {Kimble}, R.~A., {Long}, K.~S., {Blair}, W.~P., {Bowers},
  C.~W., {Dixon}, W.~V., {Durrance}, S.~T., {Feldman}, P.~D., {Ferguson},
  H.~C., {Henry}, R.~C., {Kriss}, G.~A., {Kruk}, J.~W., {Moos}, H.~W., and
  {Vancura}, O., 1992, ApJ{, 392}, 264.

\bibitem{Ketal95}
{Kruk}, J. et~al., 1995, ApJ{, 454}, L1.

\bibitem{DRO93}
{Dorman}, B., {Rood}, R.~T., and {O'Connell}, R.~W., 1993, ApJ{, 419}, 596.

\bibitem{BFD96}
{Brown}, T.~M., {Ferguson}, H.~C., and {Davidsen}, A.~F., 1996, ApJ{, 472},
  327.

\bibitem{MBWF85}
{Michaud}, G., {Bergeron}, P., {Wesemael}, F., and {Fontaine}, G., 1985, ApJ{,
  299}, 741.

\bibitem{LWF87}
{Lamontagne}, R., {Wesemael}, F., and {Fontaine}, G., 1987, ApJ{, 318}, 844.

\bibitem{BWMF88}
{Bergeron}, P., {Wesemael}, F., {Michaud}, G., and {Fontaine}, G., 1988, ApJ{,
  332}, 964.

\bibitem{KDABBBCDJKMMPWBGJNSS92}
{King}, I., {Deharveng}, J.~M., {Albrecht}, R., {Barbieri}, C., {Blades},
  J.~C., {Boksenberg}, A., {Crane}, P., {Disney}, M.~J., {Jakobsen}, P.,
  {Kamperman}, T.~M., {Macchetto}, F., {MacKay}, C., {Paresce}, F., {Weigelt},
  G., {Baxter}, D., {Greenfield}, D., {Jedrzejewski}, R., {Nota}, A., {Sparks},
  W.~B., and {Stanford}, S.~A., 1992, ApJ{, 397}, L35.

\bibitem{bertola95}
{Bertola}, F., {Bressan}, A., {Burstein}, D., {Buson}, L.~M., {Chiosi}, C., and
  {{di Serego Alighieri}}, S., 1995, ApJ{, 438}, 680.

\bibitem{SC97}
{Songaila}, A. and {Cowie}, L.~L., 1997, Hawaii Active Catalog;
  www.ifa.hawaii.edu/$\sim$cowie/tts/tts.html.

\end{thebibliography}
\bibliographystyle{nature}

%
 
\end{document}